\documentclass[12pt,aps,prl,twocolumn,showpacs,superscriptaddress,groupedaddress,footinbib,reprint,noshowpacs]{revtex4-1}
\usepackage{color}
\usepackage{graphicx}
\usepackage{amsmath}
\usepackage{xcolor}
\usepackage{comment}
\usepackage{amssymb}
\usepackage[normalem]{ulem}

\usepackage{subfigure}

\begin{document}

\title{Fluctuation-induced forces in homogeneous isotropic turbulence}
\author{Vamsi Spandan}
\affiliation{Physics of Fluids, University of Twente, Enschede 7500 AE, The Netherlands}

\author{Daniel Putt}
\affiliation{University of Houston, TX 77004, USA}
\author{Rodolfo Ostilla-M\'onico}
\email{rostilla@central.uh.edu}
\affiliation{University of Houston, TX 77004, USA}

\author{Alpha A. Lee}
\email{aal44@cam.ac.uk}
\affiliation{Cavendish Laboratory, University of Cambridge, Cambridge CB3 0HE, United Kingdom }

\begin{abstract}
Understanding force generation in non-equilibrium systems is a significant challenge in statistical physics. We uncover a surprising fluctuation-induced force between two plates immersed in homogeneous isotropic turbulence using Direct Numerical Simulation. The force is a non-monotonic function of plate separation. The mechanism of force generation reveals an intriguing analogy with fluctuation-induced forces: energy in the fluid is localised in regions of high vorticity, or ``worms'', which have a characteristic length scale. The magnitude of the force depends on the packing of worms inside the plates, with the maximal force attained when the plate separation is comparable to the characteristic worm length. A key implication of our study is that the length scale-dependent partition of energy in an active or non-equilibrium system determines force generation.
\end{abstract}

\maketitle

A fluctuating medium can exert a force on boundaries that confine the fluctuation \cite{kardar1999friction}. The most celebrated of these fluctuation-induced forces is the quantum Casimir effect: metallic plates in a vacuum experience an attractive force because they confine vacuum fluctuations of the electromagnetic field. Inspired by biological systems, recent works have focused on the force exerted by active media that continuously consume energy on confining boundaries \cite{ray2014casimir,ni2015tunable,Lee2017fluctuation}. Understanding the phenomenology of those active fluctuation-induced forces remains a theoretical challenge. A recent hypothesis suggests that the fluctuation-induced force between two plates in an active system can be a non-monotonic function of plate separation due to a non-trivial spatial partition of energy \cite{Lee2017fluctuation}, which is a prevalent feature in active fluids \cite{wensink2012meso,bratanov2015new}. Nonetheless, the force between plates in hydrodynamic turbulence, the paradigmatic non-equilibrium system with coherent spatial structures and well-studied mechanisms of energy dissipation, is hitherto unknown. 

 Interactions between objects in fluid turbulence range from relatively simple to exceedingly complex. Adequate knowledge of the interactions between objects in turbulence is necessary to understand a range a phenomena ranging from collective dynamics of planktons to volcano eruptions and multiphase flows in industrial processes. Strong interactions occur in a variety of systems, not only between the fluid and the immersed objects but also between the objects themselves. Numerical studies of both rigid spheres \cite{ten2004fully} or deformable bubbles \cite{spandan2018physical} have mainly focused on the wake instabilities of a single particle, or the collective long-range effects of swarms of particles on the underlying turbulence. Numerical studies have shown that rigid stationary objects immersed in homogeneous isotropic turbulence (HIT) affect turbulence statistics at distances more than ten times the viscous layer on the surface of the body \cite{naso2010interaction}.
 


In this Letter, we report the first direct numerical simulation (DNS) of rigid plates immersed in HIT. We demonstrate the existence of an attractive fluctuation-induced ``Casimir'' force, and reveal that the force is a non-monotonic function of plate separation. By analysing the pressure distribution {between} the objects as well as the location of the force maximum, {we show that the mechanism of force generation lies in the ability of the plates to pack specific flow features in between them, and thus causing a} Casimir-like force.

HIT \cite{ishihara2009} is a basic model for turbulence, where an idealized turbulent state is numerically simulated using a triply-periodic computational domain. The domain is randomly forced at the largest wavelengths that fit in the computational box. The energy then cascades across length-scales, energizing smaller and smaller structures up to scales where viscosity dominates (the Kolmogorov scale), and the energy is dissipated. If the forcing is continuous in time, statistically stationary states with a continuous energy spectrum can be reached. Intermediate scales are energetic, and play an active role in the cascade. They are known as ``inertial" length scales, and are neither artificially forced nor are heavily dissipative. In this inertial range there is a robust relationship between energy and wavenumber known as the ``-5/3rds law'' \cite{urielfrisch}.

\begin{figure*}
 \centering
 \label{fig:vis}
 \includegraphics[width=0.95\textwidth]{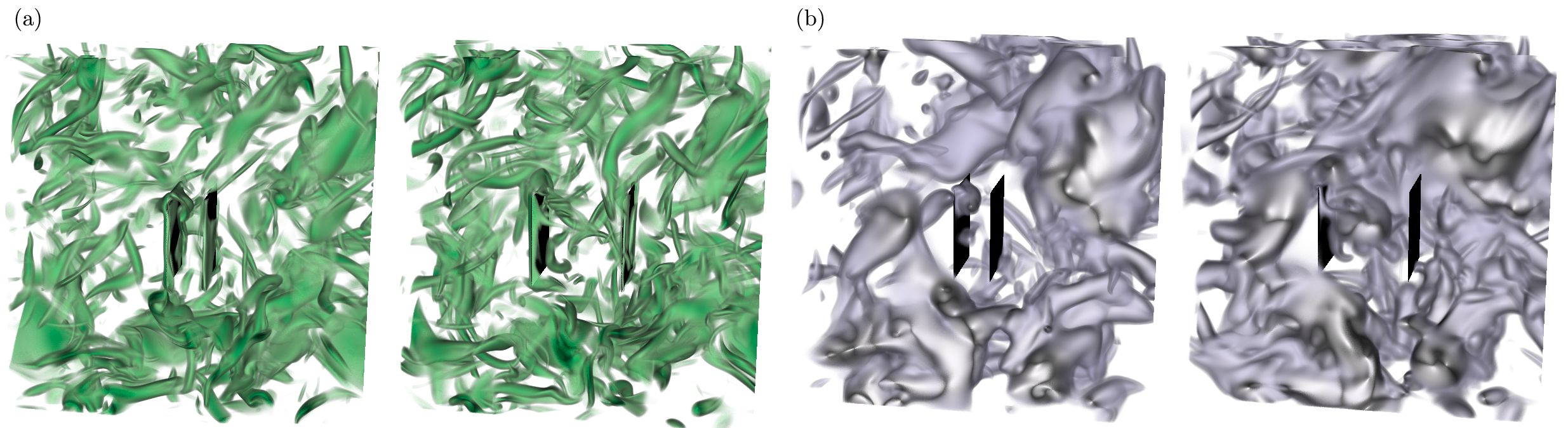}
 \caption{Turbulent structures, which carry energy, are excluded from the space in between narrowly separated plates. (a) Visualization of enstrophy (green indicates regions of high enstrophy) and (b) pressure (purple indicates regions of low pressure) for the cases with $Re_\lambda=65$, parallel square plates with separation $d=0.1\mathcal{L}$ (left panel in (a) and (b)) and $d=0.25\mathcal{L}$ (left panel in (a) and (b)). Videos are available in the supplementary material. }
\end{figure*}

When a plate is submerged in a turbulent fluid, it is subjected to two kinds of loads: viscous and pressure forces. Viscous stresses are tangential, but pressure acts in the normal direction. A difference in local pressure on both sides of a rigid plate can lead to a net force acting along its normal.
The pressure ($p$) in a fluid of density ($\rho$) is directly related to the vorticity ($\omega$), i.e.~the curl of velocity {and the rate of strain in the fluid $\sigma$} by the relation $\nabla^2(p/\rho)=0.5\omega^2-\sigma^2$ \cite{urielfrisch}. {As is shown in Figure 1}, regions of intense vorticity {tend to correlate with low-pressure regions, and organize} themselves into tubular-like structures, also known as worms \cite{jimworms}. The radius of these tubular structures scales as the Kolmogorov length $\eta_K=(\nu^3/\varepsilon)^{1/4}$ (a measure of the viscous cut-off length), and their length scales as the integral, or decorrelation length-scale, where $\nu$ is the kinematic viscosity of the fluid and $\varepsilon$ the energy dissipation in the system. An estimate for this decorrelation length-scale is provided by the large-eddy length scale, defined as $L=k^{3/2}/\varepsilon$, with $k$ the kinetic energy of the flow.

While a single plate in a randomly forced fluid will feel on average symmetric forces on both sides, by having two plates close to each other, we can restrict the {structures} that ``fit'' between the plates, thus controlling the pressure fluctuations on one side and generating a net force. This mechanism is analogous to thermal or quantum Casimir forces \cite{kardar1999friction}, where a restriction in the fluctuations imposed by the boundaries leads to force generation. Here, as the fluctuations organise themselves into a defined length scale -- the ``worm'' -- we would expect the Casimir force to be non-monotonic with the peak force achieved when the plate separation is comparable to the length scale of the worms \cite{Lee2017fluctuation}. 

To demonstrate a Casmir-like force in turbulence, we perform direct numerical simulations of HIT using the incompressible Navier-Stokes equations: 
\begin{equation}
\frac{\partial \textbf u}{\partial t}+\textbf u \cdot \nabla \textbf u=-\frac{1}{\rho_f}\nabla p+\nu\nabla^2 \textbf u +\textbf F,
\label{eqn:ns}
\end{equation}
\begin{equation}
\nabla \cdot \textbf u=0 ,
\label{eqn:con}
\end{equation}
where $\textbf u$ is the fluid velocity vector, $\rho_f$ is the density of the fluid and $p$ is the hydrodynamic pressure. $\textbf F=\textbf f^{\text{tur}} + \textbf f^{\text{ibm}}$ is a force vector which is a sum of two terms; i.e. $\textbf f^{\text{tur}}$ is a contribution from the forcing needed to generate homogeneous turbulence in the domain, while $\textbf f^{\text{ibm}}$ is the force needed to enforce the influence of the immersed plates on the flow through the immersed boundary method which is described later. The equations are solved using an energy-conserving second-order centered finite difference scheme in a Cartesian domain with fractional time-stepping. An explicit Adams-Bashforth scheme is used to discretise the non-linear terms while an implicit Crank-Nicholson scheme is used for the viscous terms \cite{verzicco1996finite}. Time integration is performed via a self starting fractional step third-order Runge-Kutta (RK3) scheme and the time-step is dynamically chosen so that the maximum Courant-Friedrich-Lewy (CFL) condition number is 1.2. The domain is periodic in all three directions with a periodicity length $\mathcal{L}$. The formulation of the force vector $\textbf f^{\text{tur}}$ is based on random processes driving the time evolution of a selected number of large scales (or small wavenumber modes). Additional details on the forcing scheme and its corresponding parameters can be found in Eswaran and Pope \cite{eswaran1988examination} and Chouippe and Uhlmann \cite{chouippe2015forcing}.

\begin{figure*}
 \centering
 \label{fig:forc}
  \includegraphics[width=0.3\textwidth]{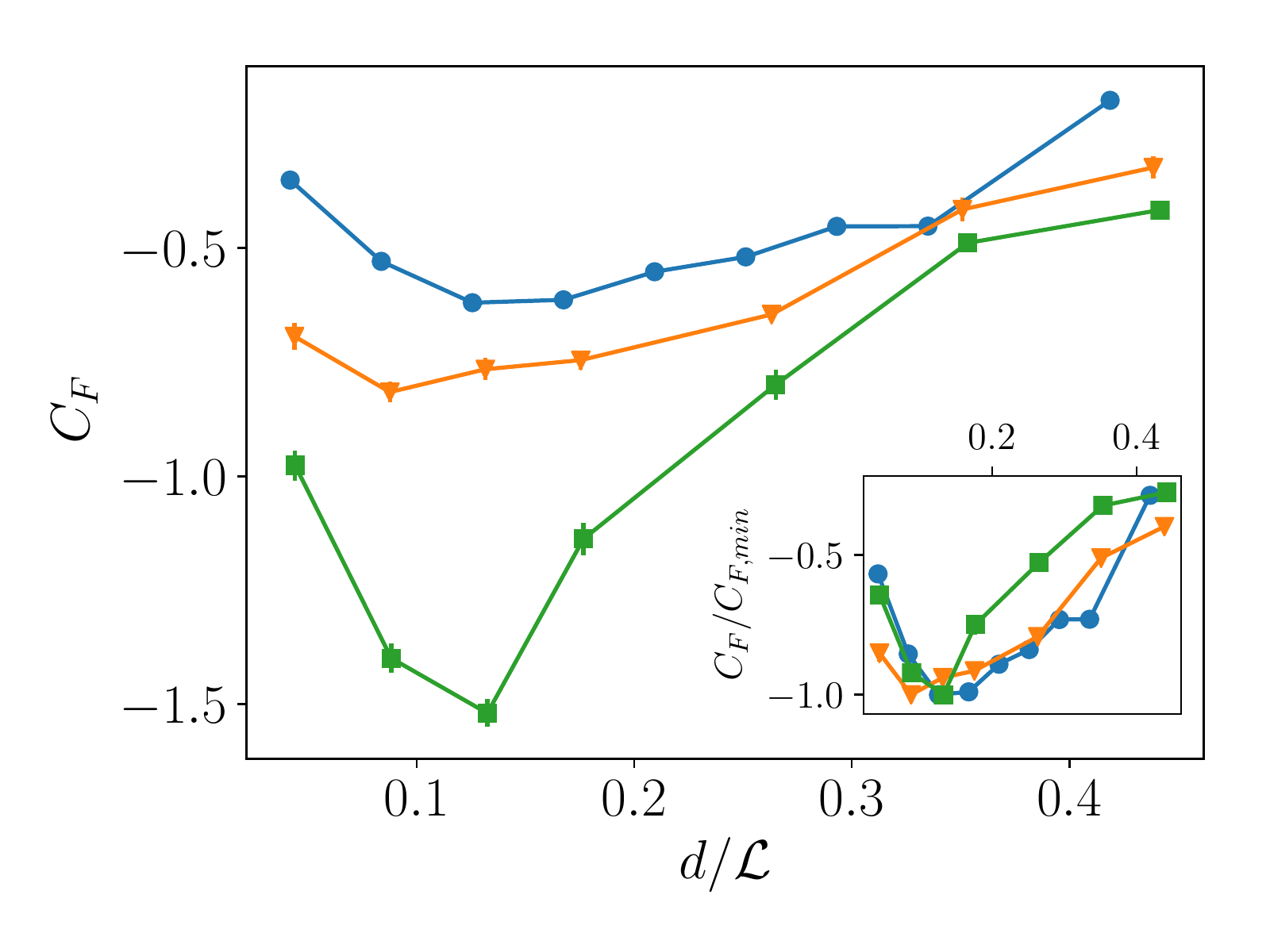}
   \includegraphics[width=0.3\textwidth]{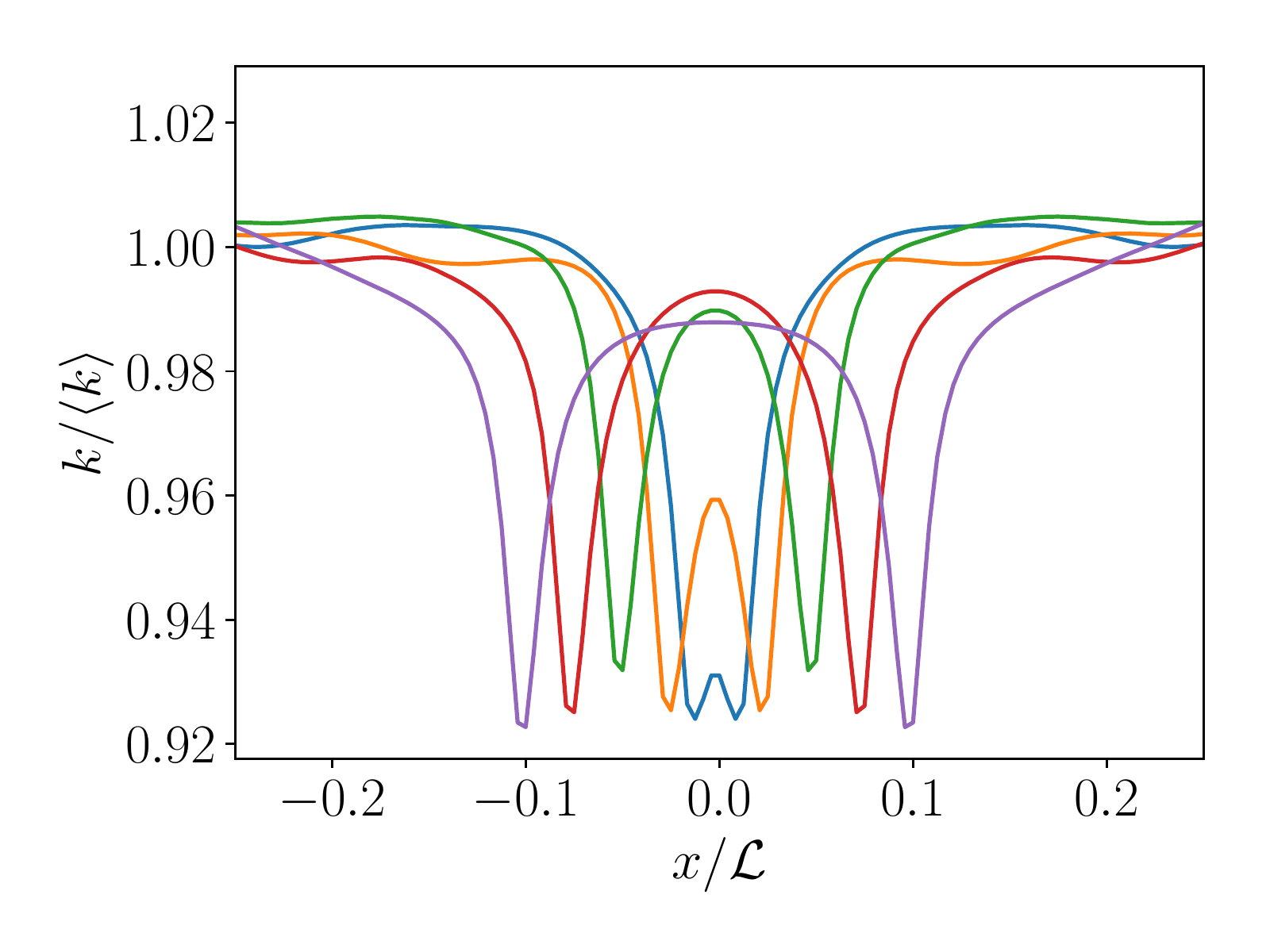}
 \includegraphics[width=0.3\textwidth]{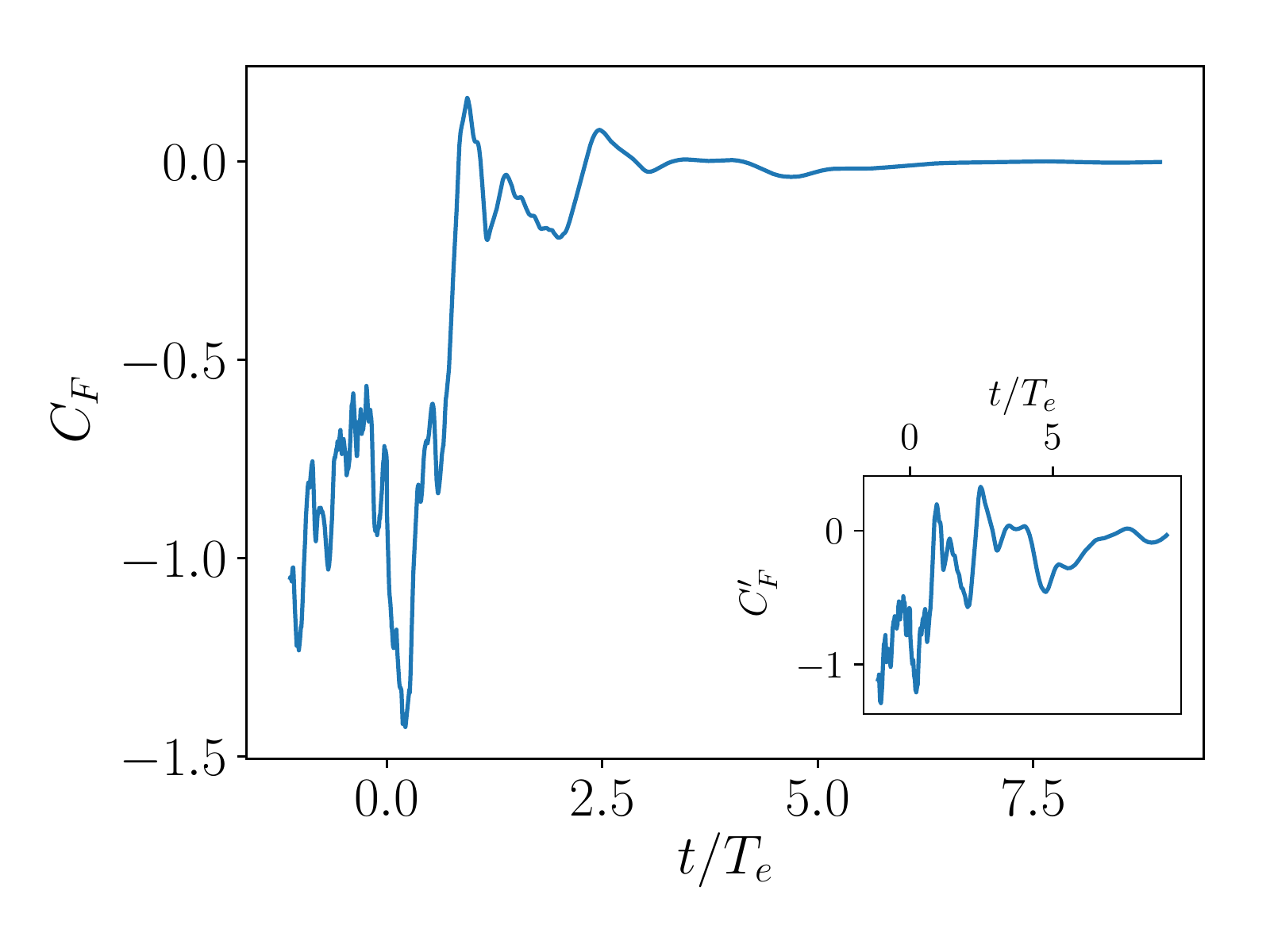}
  \caption{The turbulent Casimir force is attractive and non-monotonic, and arises as confinement modifies the fluctuation energy between the plates. Left: average attractive force on the plates with side $l_p/\mathcal{L}=0.25$ as a function of distance. The inset shows the force normalized by the value of the force minimum $C_{F,\mathrm{min}}$. The blue, orange and green solid lines are for $Re_\lambda=65$, $Re_\lambda=100$ and $Re_\lambda=140$ respectively. Middle: spatial distribution of fluid kinetic energy ($k$) in the direction normal to the plates for $Re_\lambda = 65$, $l_p/\mathcal{L}=0.25$ and several plate distances.  Right: instantaneous attractive force for unforced HIT. Forcing stops at $t/T_e=0$, where $Re_\lambda=100$ and $d/\mathcal{L}=0.25$. The main plot shows the attractive force normalized by the average $u^\prime$ for the forced case, while the inset shows the force normalized by the instantaneous (and unforced) $u^\prime$, denoted as $C_F^\prime$.}
\end{figure*}

We use a computational periodic cube of side $\mathcal{L}$, with two square parallel plates of size of $l_p/\mathcal{L}=0.25$, and vary the plate distance $d$ between $d/\mathcal{L}=0.05$ and $0.25$. Taylor-Reynolds numbers of $Re_\lambda=u^\prime \lambda/\nu=65$, $100$ and $140$ are simulated, where $u^\prime$ is the root mean squared velocity fluctuation {in one direction}, and $\lambda$ is the Taylor microscale $\lambda=\sqrt{15(\nu/\varepsilon)}u^\prime$. This $Re_\lambda$ is a measure for the range of scales covered by the inertial range. The values simulated are enough to provide a well-developed integral range while limiting computational costs. To spatially discretize the equations, a cubic grid is used, with $240^3$ points for $Re_\lambda=65$, $360^3$ for $Re_\lambda=100$, and $480^3$ points for $Re_\lambda=140$, similar to those used in \cite{chouippe2015forcing} at the same values of $Re_\lambda$. 

The influence of rigid plates on the surrounding fluid was simulated using an immersed boundary method (IBM) based on the moving least squares (MLS) approximation \cite{spandan2017parallel}. In IBM, the influence of any immersed body on a chaotic time-dependent flow environment is represented through a body-force $\textbf f^{\text{ibm}}$ in the NS equations. This force vector depends on the velocity of the immersed body (which is zero in the current simulations) and the local velocity of the fluid along the plate's surface. The interpolation and extrapolation algorithms used in the computation of $\textbf f^{\text{ibm}}$ is described in Spandan \emph{et al.} \cite{spandan2017parallel}. Since the immersed plate experiences a non-uniform distribution of local fluid velocities along the surface, the surface of the plate is discretised using approximately $2\times 10^4$ triangular elements. The force to be included in the fluid momentum equation is computed at each of these triangular elements and then distributed back on to the fluid mesh appropriately. Temporal convergence of the forces was assured by running the simulations until the force from the hydrodynamic pressure on both plates were equal (but oppositely signed) to within $3\%$. This defined the height of the errorbars in Figure 2. In practice this meant running up to more than $100$ large-eddy turnover times defined by $T_e=u^{\prime2}/\varepsilon$.

Figure 1 shows an instantaneous snapshot of the flow for two plate distances. The vorticity worms are clearly visible in figure 1a. The panels visualising enstrophy shows the effect of the plates on the worms. In the left-most panel, only the smallest of vorticity worms can penetrate between them, and when the plates are distanced further apart, similar structures exist both inside and outside the plates.

A Casimir-like force arises as the plates modify the structures between the plates by restricting the energy fluctuation modes. In order to quantify the force, {we define the force coefficient as $C_F=F/\frac{1}{2}\rho_F u^{\prime2}A$, where $F$ is the temporally averaged normal force on the plates, and $A$ their area.} Figure 2a shows that the force between the plates is attractive and is a non-monotonic function of plate separation for three different values of turbulent strengths, $Re_\lambda$. A force can arise as long as the two plates can \lq feel\rq\ each other, and this happens when the distance between them becomes comparable to the large-eddy, {or integral} length, $L$, {which is approximately constant across our simulations, $L/\mathcal{L}\in(0.569,0.597)$}. {The region between the two plates then has a significantly lower pressure than the region outside the plates due to the asymmetry in the flow structures that fit in between and outside the plates.} A non-monotonic force arises because a lower limit is set by viscosity, where the whole flow becomes quiescent between the plates and no worms {structures} fit. 

A lingering question is how do the fluctuation modes distribute themselves inside the slit and how does this contribute to the nature of the force between the plates. Figure 2(b) shows the mean kinetic energy of the fluid (averaged in the planes normal to the plates and in time) {($k=3 u^{\prime 2}/2$)} in the area between the two plates for several distances.  A narrower plate separation leads to lower mean energy inside the walls, agreeing with our physical picture that the plates partition the energy in the fluctuations in between and outside the plates. {The magnitude of the force increases as $Re_\lambda$ increases, which could indicate that the asymmetric partitioning of flow structures are responsible for the force, as it is known that the vorticity and circulation of worms increases with increasing $Re_\lambda$  \cite{jimworms}}.

{To provide further evidence that the attractive force results mainly from small-scale fluctuations and not from the large-scale structures, we consider a numerical experiment where we switch off the forcing,  i.e. $\textbf f^{\text{tur}}$ in equation (1) is set to zero, and measure the force on the plates during the decay of turbulence. The resulting force coefficient, normalized by both using the mean velocity from the forced HIT, as well as from the instantaneous decaying r.m.s. velocity is shown in Figure \ref{fig:forc}(c). We observe that after the random forcing is switched off at $t=0$, the attractive force persists for a short time interval while small-scale fluctuations still exist in the flow. After time-scales of the order of $\mathcal{O}(T_e)$, the turbulent energy cascade is drained, as there is no large-scale forcing. The large scales are unable to continue producing small-scales (whose lifetime before being damped by viscosity is $\mathcal{O}(\sqrt{\nu/\epsilon})\sim10^{-2} T_e$), and the attractive force begins to decreasing even when renormalized to account for the decaying fluctuation velocity. Large energy-containing structures have much longer life-times, estimated as $\mathcal{O}(L^2/\nu)\sim 10^3 T_e$. When $t>T_e$, i.e. when only large-scales are left in the flow, the renormalized force coefficient between the two plates is smaller in absolute value and even becomes repulsive for large times, showing that the small-scale worms are crucial for generating attractive forces. The persistence of attractive forces for time-scales of the order of $T_e$ also rules out numerical resonant effects between the random forcing and the plates. }

\begin{figure}
 \centering
 \includegraphics[width=0.4\textwidth]{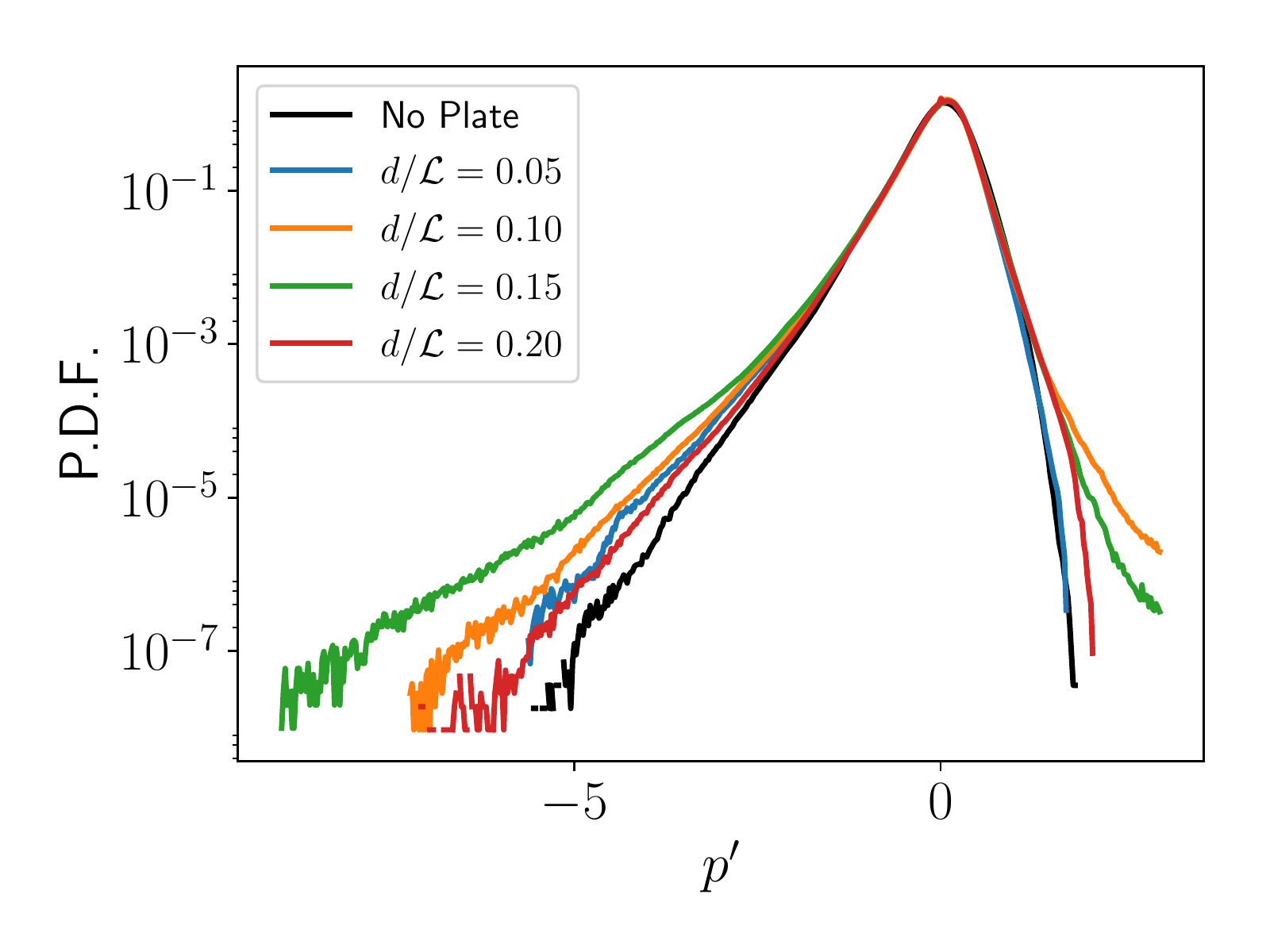}
 \caption{Pressure fluctuations for the different plate distances $d/\mathcal{L}$ for $Re_\lambda=140$. }
 \label{fig:prflu}
\end{figure}

Further physical understanding of the emergence of the fluctuation-induced force can be gained by analysing the pressure fluctuations in the flow with and without the plates. It is well known that pressure fluctuations in turbulence are asymmetric, and in particular have long tails for negative fluctuations \cite{pumirpressure}. This is due to both the intimate connection between pressure, vorticity and strain as discussed previously, and the non-Gaussian nature of turbulence. Gaussian velocity fields are known to produce very different pressure distributions in comparison to pressure distributions obtained by solving Navier-Stokes within a homogeneous-isotropic setting \cite{pumirpressure}. {The skewness in the negative fluctuations comes primarily from the intense vortex fluctuations which are a characteristic of the `worms' described previously and can be explained as a consequence of the pressure Poisson equation. Previous studies on statistics of strain conditioned on high-vorticity and vice versa in homogeneous isotropic turbulence show that high strain regions are typically associated with high vorticity while high vorticity regions are associated with weak strain regions \cite{pumirpressure}. This leads to bias in the source term for the pressure Poisson equation, i.e. $0.5\omega^2-\sigma^2$ is lower in high strain regions in comparison to high vorticity regions and since worms are characterised by intense vorticity, their presence leads to a skewness in the negative pressure fluctuations. By immersing plates separated by a given distance into the flow, we selectively filter the flow fields and consequently strengthen the intense negative pressure fluctuations by partitioning the intense vorticity regions. This is shown in Figure \ref{fig:prflu} which shows a probability distribution function (P.D.F.) of the pressure fluctuations for $Re_\lambda=65$ for all the simulated plate separations in the entire domain. While the tails of the distribution are affected for both negative and positive values of $p^\prime$, for plate separations with a significant Casimir-like force, the tails of the pressure fluctuation P.D.F. is significantly wider, with extreme values of negative pressure, indicating that the fluctuation modes are modified by the presence of the plates which leads to an attractive force certain plate separations. }

In summary, we demonstrate the existence of an attractive turbulent Casimir-like force by direct numerical simulation of two plates immersed in an homogeneous isotropic turbulent flow. Our study sheds light on the interaction between objects in a turbulent flow at very close distances and can be relevant to collective sedimentation processes in a turbulent environment. {We anticipate experimental investigations of this force in a von Karman setup which has been previously used for fundamental studies in single-phase and multi-phase homogeneous isotropic turbulence \cite{brouzet2014flexible}.} Moreover, we speculate that this mechanism of generating a non-monotonic Casimir-like force due to non-trivial spatial partition of energy might be a general phenomenology for active and non-equilibrium systems. {In particular, recent pioneering works show that a large class of biological fluids comprising microbial suspensions exhibit striking analogies with turbulent flows (e.g. the velocity field is chaotic and intermittent, and the energy spectra display power law behaviour), and ``active turbulence'' can be modelled with a Navier-Stokes equation with higher order gradient terms in the stress tensor \cite{wensink2012meso,bratanov2015new,kokot2017active,mickelin2018anomalous}. The novel mechanism of force generation that we reported between objects in a classical turbulent flow opens up the possibility that similar mechanisms might exist in more exotic turbulent-like systems. Forces between objects in active turbulence is relevant to understanding the organisation in biofilms with dead and active bacteria \cite{chai2011extracellular,drescher2011fluid}, using active fluids as a medium to control self-assembly \cite{angelani2011effective}, and using active fluids to power micromachines \cite{thampi2016active}. }

\begin{acknowledgements}
This work was completed in part with resources provided by the University of Houston Center for Advanced Computing and Data Science. AAL acknowledges the support of the Winton Programme for the Physics of Sustainability. The authors thank R. Verzicco, A. Pumir, Y. B. Sinai and M. P. Brenner for valuable discussions. 
\end{acknowledgements}

\bibliography{refs.bib}

\end{document}